\documentclass[12pt]{article}
\usepackage{epsfig}

\def\beq{\begin{equation}}
\def\eeq{\end{equation}}
\def\bea{\begin{eqnarray}}
\def\eea{\end{eqnarray}}
\def\noi{\noindent}

\begin{document}

\rightline{October 2001}
\rightline{UCOFIS 3/01}
\rightline{US-FT/5-01}
\rightline{LPT Orsay 01-39}
\vspace{0.5cm}

\begin{center}
{\Large\bf Analysis of the first RHIC results in the String Fusion
Model}
\vspace{0.5cm}

N. Armesto$^{a,\dag}$,
C. Pajares$^{b,\ddag}$
and
D. Sousa$^{c,\S}$

\vspace{0.2cm}

$^a$ {\it Departamento de F\'{\i}sica, M\'odulo C2, Planta baja, Campus
de Rabanales,}\\
{\it Universidad de C\'ordoba, E-14071 C\'ordoba, Spain} \\

$^b$ {\it Departamento de F\'{\i}sica de Part\'{\i}culas, Universidade
de Santiago de Compostela,}\\
{\it E-15706 Santiago de Compostela, Spain}\\

$^c$ {\it Laboratoire de Physique Th\'eorique, Universit\'e de Paris
XI,}\\
{\it B$\hat{a}$timent 210, F-91405 Orsay Cedex, France}\\

\end{center}

\noi{\footnotesize $^\dag$ E-mail: fa1arpen@uco.es
\hskip 3cm $^\ddag$ E-mail: pajares@fpaxp1.usc.es}

\noi{\footnotesize $^\S$ E-mail:
Dolores.Sousa@th.u-psud.fr}

\vfill
{\small
\centerline{\bf Abstract}
First results from RHIC on charged multiplicities, evolution of
multiplicities with centrality, particle ratios and transverse momentum
distributions in central and minimum bias collisions,
are analyzed in a string model which includes hard collisions,
collectivity in the initial state considered as string fusion, and
rescattering of the produced secondaries. Multiplicities and their
evolution with centrality are successfully reproduced. Transverse
momentum distributions in the model show a larger $p_T$-tail than experimental
data, disagreement which grows
with increasing centrality. Discrepancies with particle
ratios appear and are examined comparing with
previous features of
the model at SPS.
}

\newpage

With the first collisions at the Relativistic Heavy Ion Collider (RHIC)
at BNL in June 2000, the study of nuclear collisions has entered the
truly ultrarelativistic domain. While there exist predictions from many models
\cite{lastcall}, now experiments have presented results
\cite{phobosprl,starprl,phenixprl,phenixprl2,brahms,
phenix,phobos,star,phobosratios,starratios,brahmsratios,newphobos} on
several aspects of data, most of them corresponding to AuAu collisions at 130
GeV per nucleon in the center of mass. So it comes the time to examine the
ability of models for ultrarelativistic heavy ion collisions,
fitted to describe nuclear data at the much
lower energies of the Super Proton Synchrotron (SPS) at CERN and nucleon
data in the range of energies going from SPS to TeVatron at FNAL,
to describe the
new situation, and whether the evidences of Quark Gluon Plasma (QGP)
already obtained at SPS are verified or not \cite{qgpan}. The aim of this
letter is to compare the results of the String Fusion Model (SFM) 
\cite{sfm,proximo} with some of
the first RHIC data. Other comparisons can be found in
\cite{kari,zabrodin}\footnote{In \protect{\cite{zabrodin}} a model which, like
ours, contains multipomeron exchange, a hard component and rescattering
of secondaries, but no string fusion, is shown to be able to
reproduce the experimental data
\protect{\cite{starprl}} on elliptic flow.}.
After a very brief model description, charged
multiplicities at midpseudorapidity in central collisions,
evolution of charged multiplicities at midpseudorapidity with centrality,
transverse
momentum distributions of charged particles at different centralities
and ratios of different particles will be compared with available data
coming from the experiments. Finally some conclusions will be
summarized.

An exhaustive description of the model can be found in \cite{proximo}.
Its
main features are the following: Elementary inelastic collisions (binary
nucleon-nucleon collisions) are considered
as collisions
between partons from nucleons of
the projectile and the target, distributed in the
transverse plane of the global
collision. Some of these elementary collisions are taken as hard
ones, and proceed as gluon-gluon $\longrightarrow$ gluon-gluon
through PYTHIA \cite{pyth}
with GRV 94 LO parton density functions (pdf's) \cite{grv94} and EKS98
modification of pdf's
inside nuclei \cite{eks},
with subsequent
radiation and fragmentation performed by ARIADNE \cite{ariad}
and JETSET \cite{pyth}. Those collisions not being considered hard produce soft
strings in pairs.
These strings are allowed to fuse if their parent partons are close
enough in impact parameter \cite{sfm}; as the number of strings
increases with increasing energy, atomic number and centrality, this
mechanism accordingly grows in importance. Fragmentation of soft strings is
performed using the
tunneling mechanism
for mass and transverse
momentum distributions, while longitudinal momenta are simulated
by an invariant area law. The main consequences of string fusion are a
reduction of multiplicities in the central rapidity region and an
increase in heavy particle production.
The produced particles are allowed to
rescatter (between themselves and with spectators nucleons)
using a very naive model with no proper space-time evolution,
whose consequences are a small multiplicity reduction, an increase in
strange and multistrange baryons and nucleon annihilation. Some
comments are in order at this point:
First, partons which generate both soft and hard strings can be valence quarks
and diquarks,
and sea quarks and antiquarks, so the number of soft strings is not
simply proportional to the number of wounded nucleons but has some
proportionality, increasing with increasing energy, centrality
and nuclear size, on the
number of binary nucleon-nucleon collisions\footnote{Usually the soft
contribution is taken as proportional to the number of wounded nucleons,
while the contribution proportional to the number of binary
nucleon-nucleon collisions is considered hard. Let us stress that this is
a misleading (model dependent)
statement: some proportionality with the number of binary
nucleon-nucleon collisions is demanded by a basic requirement of the
theory as unitarity, and has nothing to do with the soft or hard origin 
of these binary nucleon-nucleon collisions.}.
Besides, only fusion of two strings
in considered in the actual version of the model, and hard strings are not
fused. Finally, the rescattering model is simplistic and has been
included just to estimate the effects that such kind of physics could have and
to tune the parameters of the model as an initial condition for a more
sophisticated evolution; thus, results depending strongly on it should be taken
with great
caution. All these aspects will be commented more extensively when the
comparison with experimental data is performed.

\begin{figure}[htb]
\begin{center}
\epsfig{file=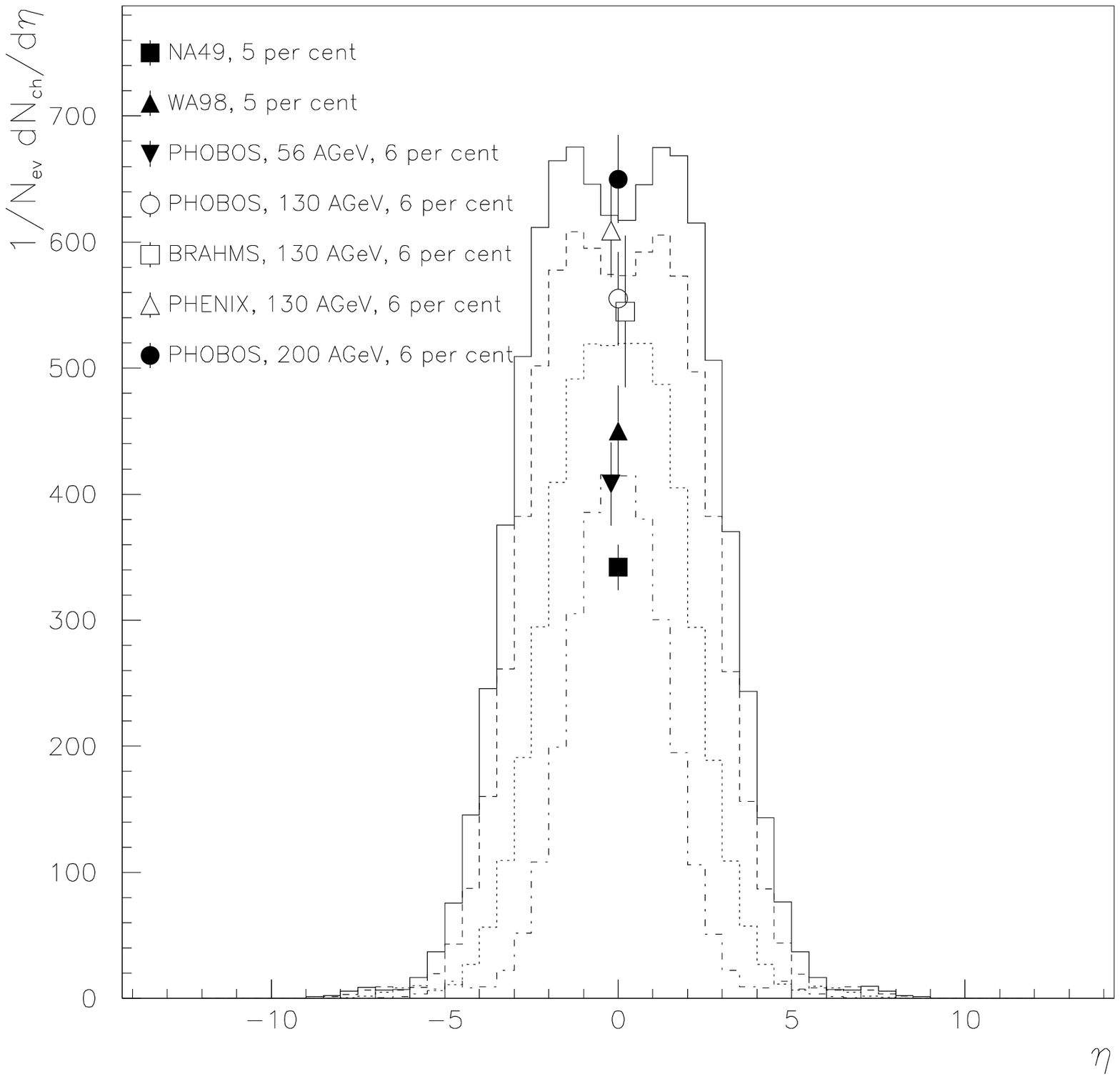,width=15.5cm}
\end{center}
\vskip -1.0cm
\caption{}
\label{fig1}
\end{figure}

In Fig. \ref{fig1} results of the model (unless otherwise
stated, results of the model correspond to its default version with
the mentioned pdf's and string fusion and rescattering, see \cite{proximo})
for the pseudorapidity distribution of
charged particles in central collisions at SPS and RHIC are compared
with experimental data. For central AuAu collisions at 130 and 200 GeV per
nucleon in the center of mass, the model successfully reproduce the
data (the ratio of multiplicities at 200 and 130 GeV is 1.08 in the
model, slightly smaller than the experimental value 1.14$\pm$0.05
measured by
PHOBOS \cite{newphobos}), while at 56 GeV it overestimates the PHOBOS results
\cite{phobosprl}. Nevertheless, the situation at these energies is not
clear: WA98 results \cite{wa98cent} at SPS lie above the PHOBOS data at
56 GeV, and far above NA49 data \cite{na49mult} (as extracted in
\cite{phobosprl}) at SPS; NA49 results on multiplicities in
central PbPb collisions at SPS are in agreement with those from WA97
\cite{wa97mult}. So it is difficult to conclude anything
definitive on the evolution from SPS to RHIC, of multiplicities with increasing
energy in the model.

\begin{figure}[htb]
\begin{center}
\epsfig{file=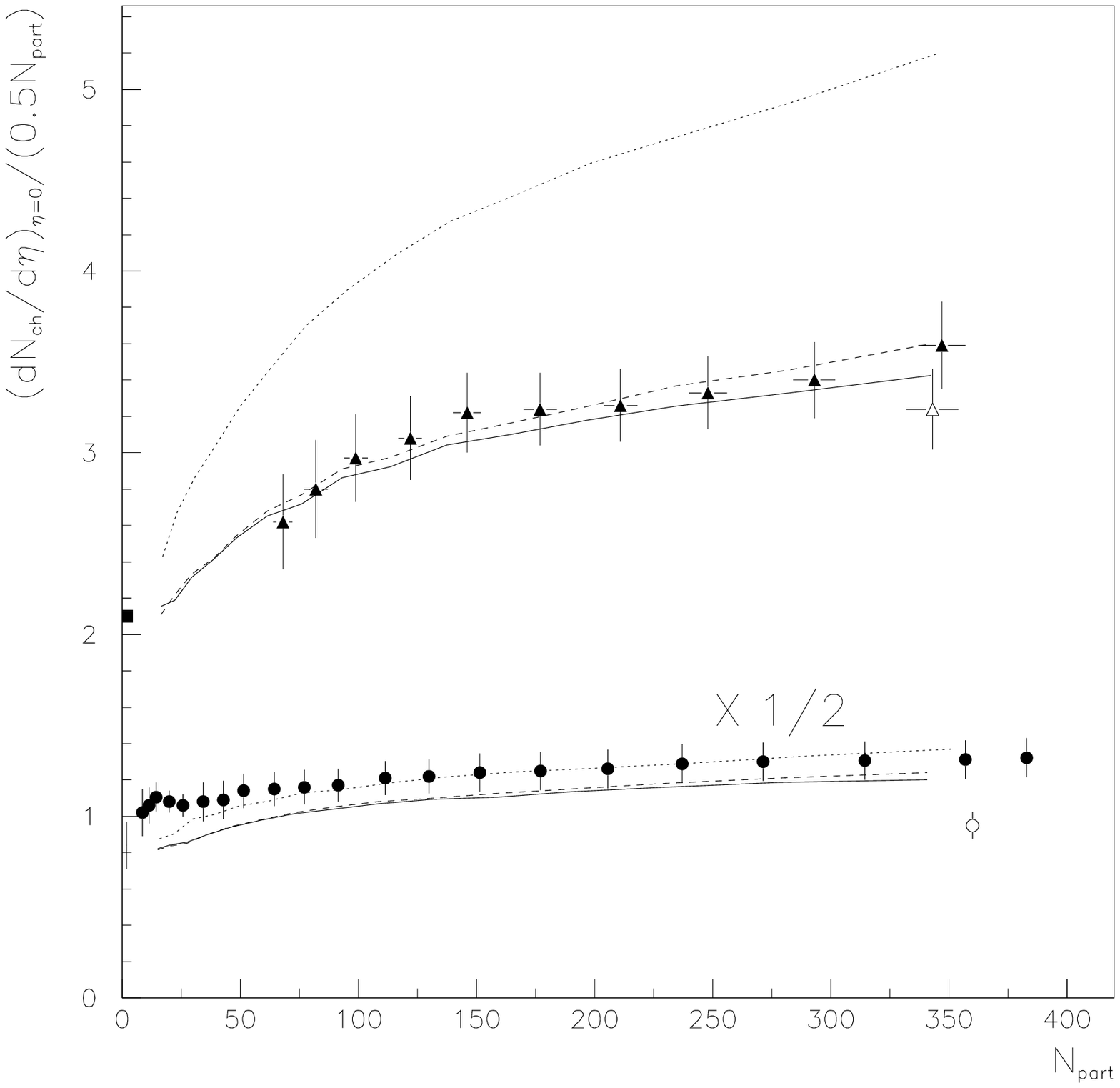,width=15.5cm}
\end{center}
\vskip -1.0cm
\caption{}
\label{fig2}
\end{figure}

\begin{figure}[htb]
\begin{center}
\epsfig{file=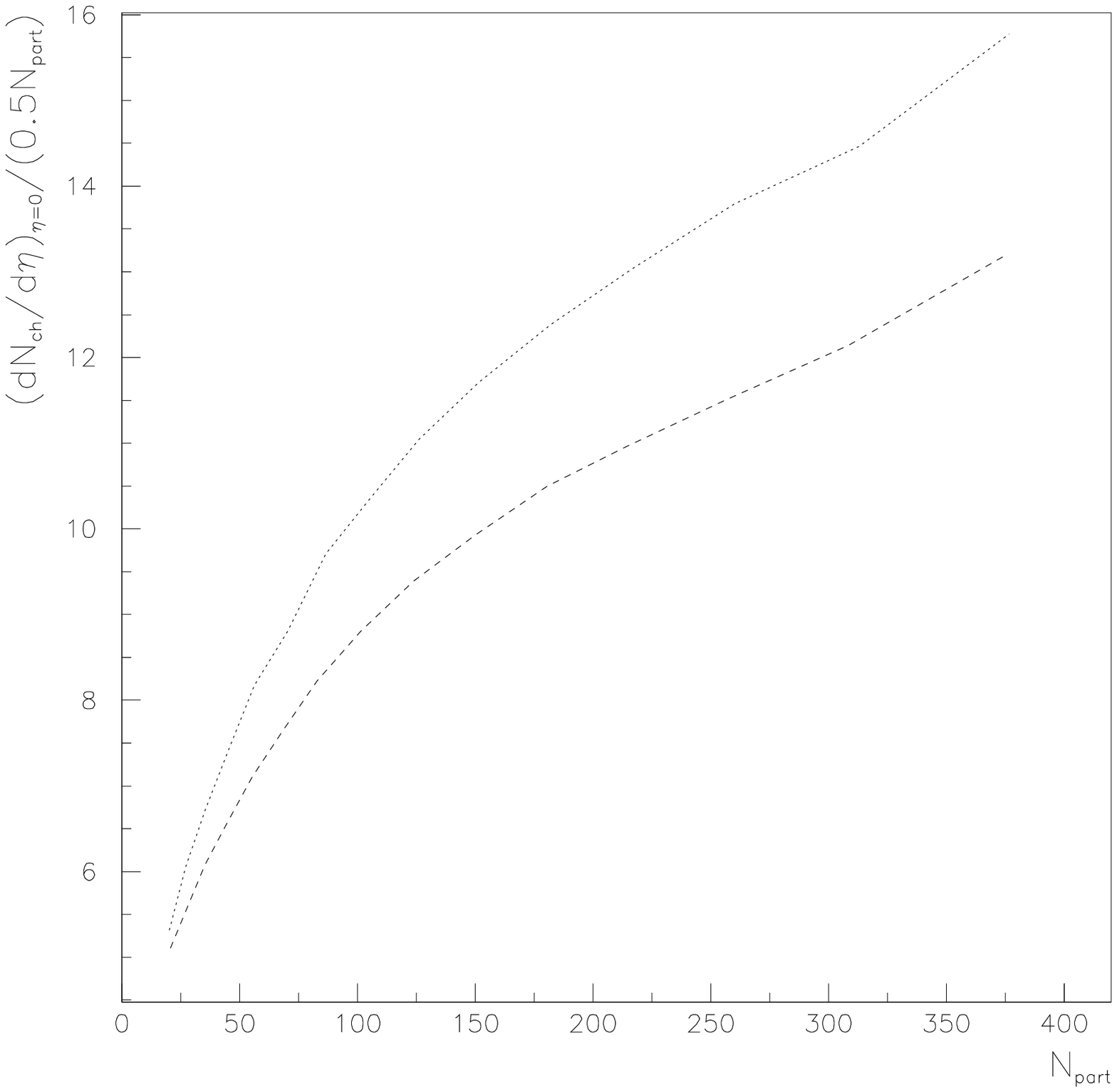,width=15.5cm}
\end{center}
\vskip -1.0cm
\caption{}
\label{fig3}
\end{figure}

Recently it has been proposed \cite{guywang} that the evolution of
multiplicities with centrality can be used as a tool to discriminate
among several models for multiparticle production in high-energy nuclear
collisions. In this way, models
which consider saturation \cite{satur1}
of either the number of partons in the wave
function of the projectile and target or in the number of partons
produced in the collision \cite{satur2}, show a constant or slightly
decreasing behavior of the
multiplicity per participant (wounded) nucleon with increasing number of
participants\footnote{Other proposals which include saturation
\protect{\cite{dima}} show an increasing behavior compatible with
data.}. On the other
hand, models which consider some proportionality with the number of
binary nucleon-nucleon collisions based on the AGK cancellation
\cite{abra}, being this proportionality already present in the soft
component \cite{proximo,alfons,triple2,dias} or only in the hard
component \cite{hijing}, show a behavior,
with the multiplicity per participant increasing with increasing number
of
participants,
qualitatively or quantitatively compatible with data. The results of our
model for the 75 \% more central collisions
at SPS and RHIC are shown in Fig. \ref{fig2} and compared with
experimental data. It can be seen that the model underestimates WA98
data at SPS, while it overestimates those from NA49, as could be
expected from the discussion about Fig. \ref{fig1}, but the qualitative
behavior seems correct. At RHIC the
agreement with data is quite satisfactory. It can be seen that the
inclusion of rescattering results in a slight decrease of
multiplicities, while the influence of string fusion is relatively small
at SPS but very important at RHIC and crucial for the agreement with
experimental data. In our model it is this latter mechanism the one
which plays the
r$\hat{\rm o}$le
of shadowing corrections in \cite{alfons,triple2,hijing}, parton
saturation in \cite{satur2,dima} or string percolation \cite{percol1}
in \cite{dias}. Concerning
the limitation of fusion of just soft strings in groups of two,
let us point out that it
seems to be compensated at
RHIC
with the choice of the fusion strength, while the non-inclusion of
fusion of hard strings is unimportant, as they amount for just 1 \% of
the total number of elementary
inelastic collisions. This is no longer the case for the future Large
Hadron Collider (LHC) at CERN,
situation for which we present the results of the model in Fig.
\ref{fig3} (results with rescattering are not presented because
this mechanism is too CPU-time consuming at LHC energies for large
nuclei):
Here, the fusion of just two strings has reached its limit,
so multiplicities are not so strongly damped as at RHIC, and fusion of
more than two strings (and of hard strings, which now amount for
32 \% of the total number of elementary inelastic collisions),
or even a phase transition
like percolation \cite{percol1}, have to be introduced in the
model.

\begin{figure}[htb]
\begin{center}
\epsfig{file=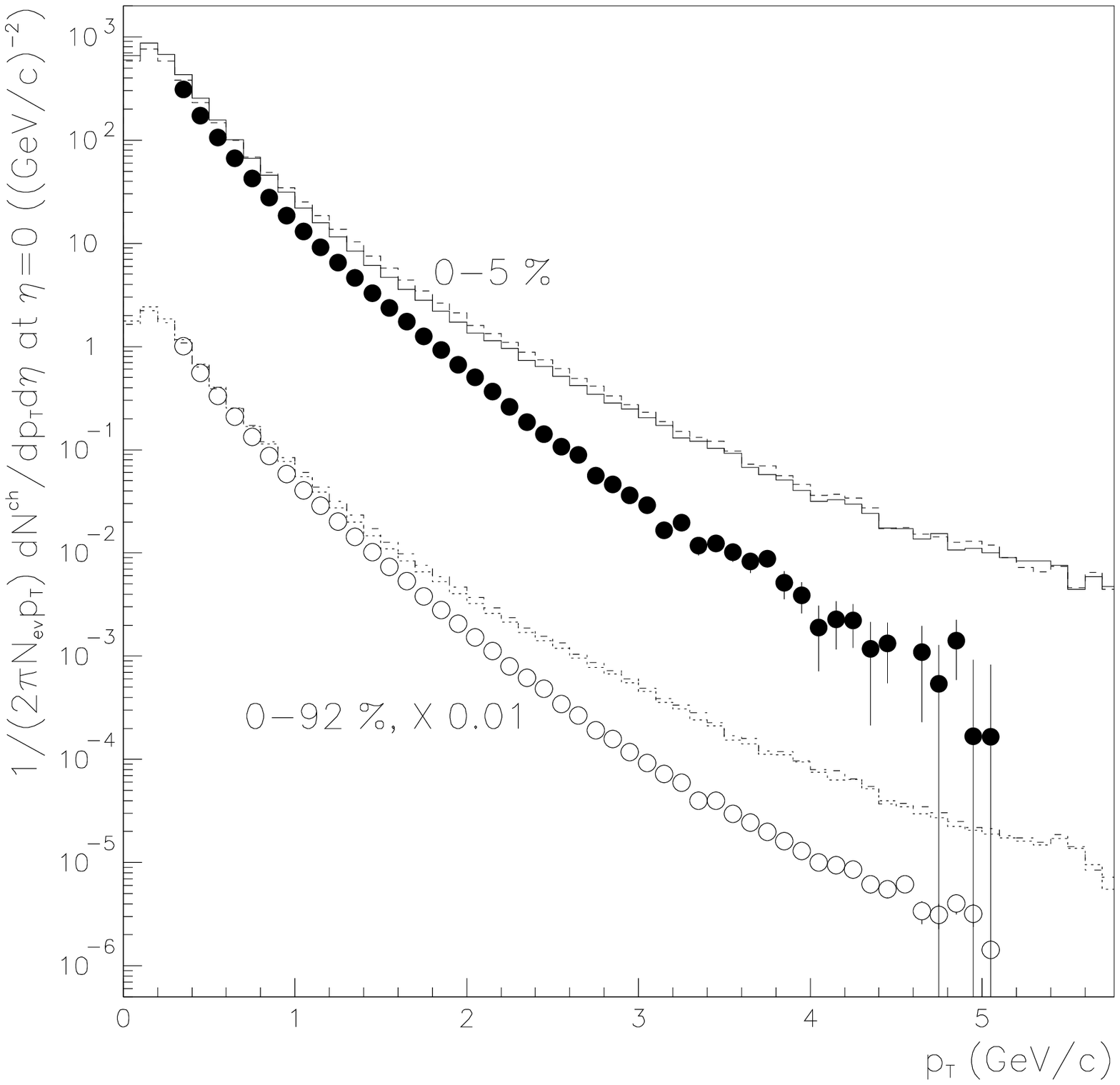,width=15.5cm}
\end{center}
\vskip -1.0cm
\caption{}
\label{fig4}
\end{figure}

Let us now turn to the transverse momentum spectrum. Preliminary
measurements \cite{phenix,star} show that the spectrum in AuAu
collisions at 130 GeV per nucleon in the center of mass falls with
increasing $p_T$ faster than predictions from models
\cite{hijing} which
reproduce the $p_T$-distributions in $\bar{\rm p}$p collisions at 200 GeV in
the center of mass; this discrepancy grows with increasing centrality.
A possible explanation is jet quenching
\cite{quenching1}, i.e. the
energy loss of high energy partons in a hot medium containing free color
charges. So, there has been a great debate on the explanation of the
absence of jet quenching at SPS and its presence at
RHIC \cite{quenchingqm}, and its interpretation as a QGP signature.
In our model
we find quite the same feature as in \cite{hijing}, see
Fig. \ref{fig4}, namely an excess of particles with high $p_T$ compared with
experimental data, excess which becomes less pronounced when going from
central to minimum bias collisions. Our model correctly reproduces
multiplicities and their evolution with centrality
at this energy (as seen in Figs. \ref{fig1} and
\ref{fig2}), and the $p_T$-spectrum in pp collisions at SPS and in
$\bar{\rm p}$p collisions at S$\bar{\rm p}$pS at CERN and TeVatron, and the
increase of $\langle p_T\rangle$ with energy and multiplicity (see
\cite{proximo}); we have
also checked that this is neither an effect of
pdf's or of their nuclear modifications, nor of
rescattering, whose influence on the $p_T$-spectrum is tiny,
see \cite{proximo} and Fig. \ref{fig4}; in fact, from the studies in
\cite{proximo} it can be concluded that the transverse momentum enhancement in
collisions between nuclei compared to those between nucleons
is due in the model both to the hard contribution which becomes more important
with an increasing number of elementary collisions, and, above all,
to the transverse
momentum broadening of the partons at the ends of the strings
introduced in the model and responsible of the increase of $\langle p_T\rangle$
with increasing multiplicity, while string fusion has a very small effect.
It is also remarkable that the discrepancy with the experimental data
appears in a model
like ours, which for the collisions studied
at RHIC produces only 1 \% of hard elementary collisions, and in a model
like that of \cite{hijing}, in which most of particle production at RHIC
energies comes from the hard contribution\footnote{Possible differences
in the $p_T$-spectrum in nucleon-nucleon collisions between our model
and those based on hard scatterings like HIJING \protect{\cite{hijing}}
should become visible at LHC, where the results are not so tightly
constrained by the existing experimental data at SPS, S$\bar{\rm p}$pS
and TeVatron: In our model the contribution from hard scatterings will be
smaller and thus we expect less high-$p_T$ particles.}. So it really
looks like an
effect which diminishes the number of high $p_T$ partons, leading them to
the low $p_T$ region. Jet quenching
\cite{quenching1,quenchingqm} seems a good candidate to
explain
this
experimental finding, but it should be taken into account that
it also leads to the appearance of more
particles at low $p_T$ and $\eta$; thus, the simultaneous comparison of
the evolution of
both multiplicities and transverse momentum distributions with
centrality should be
a crucial test for this mechanism\footnote{In \protect{\cite{vitev}} the
evolution of $\langle \bar {\rm p}\rangle/\langle \pi^- \rangle$ versus
$p_T$ with centrality
is proposed as a test of jet quenching;
the increase of this ratio with
increasing $p_T$ observed by PHENIX \protect{\cite{phenix}} is reproduced
with a soft exponential component proportional to the
number of participants plus a quenched perturbative distribution
proportional to the number of binary collisions. In our model,
the corresponding
increase due to the soft
part would be stronger than in \protect{\cite{vitev}} due to string
fusion and to the fact that this component is, in our case,
proportional to the number of both wounded nucleons and binary collisions.}.
One would think that the presence of saturation of
low transverse momentum partons \cite{satur1,satur2} would make the
comparison with experimental data even worse: the low $p_T$ region of
the spectrum, populated of poorly resolved partons, would be damped
due to parton fusion and the spectrum become flatter than without
saturation. Quite the same would occur in percolation of strings
\cite{percol1}: soft strings have a larger transverse dimension
than hard partons and would fuse more easily, and fused strings with higher
string tension would produce particles with higher $p_T$ than ordinary
strings, so the mean $p_T$ would increase with atomic size or centrality
\cite{percolpt}, contrary to what data apparently show\footnote{A recent
analysis \protect{\cite{newperc}} shows that
nevertheless it is
possible to simultaneously explain the evolution with centrality
of both multiplicity
distributions and transverse momentum spectra in a very crude realization of
the percolating string
approach.}.

Finally, in Table \ref{tab1} model results for different particle
ratios are shown and compared with published experimental data
\cite{brahmsratios,nuxu,brahms,phenix,star,phobosratios,starratios}.
For completeness, let us indicate the results in the model for
the ratios $\bar \Lambda /\Lambda$, $\bar
\Xi^+/\Xi^-$, $K^+/K^-$, $\bar {\rm p}/\pi^-$ and $K^-/\pi^-$ at $\eta\sim 0$,
for which we get 0.85$|$0.87$|$0.87, 0.60$|$0.92$|$0.88,
1.08$|$1.03$|$1.04, 0.02$|$0.07$|$0.04
and 0.08$|$0.12$|$0.16
respectively without string fusion or rescattering$|$with string fusion$|$with
string fusion and rescattering\footnote{These
results can be
compared with preliminary, not yet published results:
0.73$\pm$0.03, 0.82$\pm$0.08,
1.12$\pm$0.01$\pm$0.06, 0.08 and 0.15 respectively,
presented by STAR at QM2001 \protect{\cite{star}}.}.
The results in the model have been obtained in
the corresponding pseudorapidity regions, for AuAu collisions at 130 GeV
per nucleon in the center of mass with a centrality of 10 \% and for
particles with $p_T> 0.2$ GeV/c. Each experiment applies different
centrality and kinematical cuts for the different ratios, but a common
conclusion of all of them is that ratios are very weakly dependent on
centrality of the collision
and $p_T$ of the particles, so this should not seriously affect the
comparison. From these results
it can be seen that the model overestimates
antibaryon production, a feature already present at SPS, see \cite{proximo},
but string fusion is needed to increase the strangeness and
antibaryon yield, which is badly
underestimated, see the comparison with SPS data in \cite{proximo},
if this mechanism is not included (in the ratios at central rapidities
and due to the
lack of stopping at RHIC energies, see below, and to the fact that
string fusion creates on average the same amount of baryons and
antibaryons, this feature is mainly visible in
those involving multistrange baryons or in $\bar {\rm p}/\pi^-$).
This discrepancy is less pronounced for $\Xi$'s than for $\Lambda$'s,
and for $\Lambda$'s than for nucleons, and is more pronounced in the
central region of (pseudo)rapidity.
As stated in the brief model description, our
rescattering model is simplistic, and cannot be expected to produce
correct quantitative results, only the trend which it shows
should be considered. So all that we can conclude is that for the ratios
at RHIC, similar problems appear than those already present at
SPS\footnote{Apparently, the antibaryon-to-baryon ratios measured at
RHIC favor \protect{\cite{star}} a
coalescence model \protect{\cite{lastcall,coal}}, see
\protect{\cite{proximo}} for a comparison of our results at full RHIC
energy and
those coming from other models.}.
As a last comment, a preliminary, non-corrected for hyperon decay,
measurement of the p--$\bar{\rm p}$ yield at midpseudorapidity by BRAHMS
\cite{brahms}, gives $8\div 10$ for a centrality of 6 \% (a value $4\div
6$ has been extracted \cite{nuxu} from preliminary STAR data for the
same centrality), while in our
model we get a lower value $\sim 2$; this may suggest that the problem
in the $\bar {\rm p}$/p ratio lies not only in a $\bar {\rm p}$ excess,
but also in some lack of stopping in the model.

In conclusion, we have compared the results of the SFM with some of the
first RHIC data. At RHIC, charged multiplicities in the central region for
central collisions and their evolution with centrality are
successfully reproduced, suggesting the presence of some mechanism, like
string fusion, which
moderates the increase of multiplicities with increasing centrality;
On the other hand and in view of the SPS data, it is difficult to obtain clear
conclusions from the behavior of multiplicities in
the transition from SPS to RHIC.
Results on particle ratios show,
when compared to experimental data, similar problems of antibaryon
excess previously found at SPS, and are probably related to the
oversimplification of the model of rescattering and to problems with data
at SPS, see \cite{proximo}. Finally, in the SFM the
$p_T$-spectrum at RHIC is flatter than in data and this problem gets worse with
increasing centrality, a feature which also appears in
other models \cite{hijing,quenchingqm} in which the contribution of hard
elementary collisions is much larger than in ours. At first sight, it
looks improbable
that parton saturation or percolation of strings could improve the
comparison with the $p_T$-distributions (but see \cite{newperc}).
So, from our point of
view these data are most striking and, if confirmed,
maybe a good candidate for a signature of
non-conventional physics appearing in heavy ion collisions at RHIC.
Although the results of the model
on features which should depend strongly on the
evolution of the system (particle ratios and $p_T$-spectrum if jet
quenching is present) cannot be
considered satisfactory, the agreement with multiplicities and their
evolution with centrality, which are usually assumed not to vary too
much during evolution \cite{satur2,dima}, gives us some confidence in the
ability of the model to describe the initial condition, to be used for
further evolution, in a collision
between heavy ions at high energies.

\vskip 1cm
\noi {\bf Acknowledgments:}
We express our gratitude to A. Capella, K. J. Eskola and R. Ugoccioni
for useful discussions, and to F. Messer for providing us the
preliminary PHENIX data in Fig. \ref{fig4}.
N. A. and C. P.
acknowledge financial support by CICYT of Spain under contract
AEN99-0589-C02.
N. A. and D. S. also thank Universidad de
C\'ordoba and Fundaci\'on Barri\'e de la Maza of Spain respectively, for
financial support.
N. A. thanks Departamento de F\'{\i}sica de
Part\'{\i}culas of the Universidade de Santiago de Compostela
for stays during which part of this work was completed.
Laboratoire de Physique Th\'eorique is Unit\'e Mixte de Recherche --
CNRS
-- UMR n$^{\rm o}$ 8627.


\section*{List of figures}

\noi {\bf 1.} Results of the model for the pseudorapidity distribution of
charged particles for central (5 \%) PbPb collisions at
17.3 GeV per nucleon in the center of mass
(dashed-dotted line), and central (6 \%) AuAu collisions at 56
(dotted line),
130 (dashed line) and 200 (solid
line) GeV per nucleon in the center of mass, compared
with experimental data at SPS from NA49 \cite{na49mult,phobosprl}
(black square) and WA98 \cite{wa98cent} (black, upward pointing triangle), and
at RHIC from PHOBOS \cite{phobosprl,newphobos}
(black, downward pointing triangle for
56 GeV, open circle for 130 GeV and black circle for 200 GeV),
BRAHMS \cite{brahms} (open square) and
PHENIX \cite{phenixprl} (open triangle). 

\vskip 0.5cm
\noi {\bf 2.} Pseudorapidity density of charged particles at $\eta=0$
divided by one half the number of participant nucleons, versus the
number of participant nucleons, in PbPb collisions at 17.3 GeV per
nucleon in the center of mass (multiplied by 1/2, lower curves and
symbols) and in AuAu collisions at 130 GeV per nucleon in the center of
mass (upper curves and symbols); also the experimental
number for $\bar{\rm p}$p
collisions at 130 GeV per nucleon is given \cite{ua5}, filled square.
Experimental data are from PHENIX \cite{phenixprl} (filled triangles),
PHOBOS \cite{phobosprl} (open triangle), WA98 \cite{wa98cent} (filled
circles) and NA49 \cite{na49mult,phobosprl} (open circle). Curves are
results of the model for the 75 \% more central events, without fusion
or rescattering (dotted lines), with fusion (dashed lines) and with fusion and
rescattering (solid lines).

\vskip 0.5cm
\noi {\bf 3.} The same as in Fig. \ref{fig2}, but for PbPb collisions at 5.5 TeV
per nucleon in the center of mass.

\vskip 0.5cm
\noi {\bf 4.} Transverse momentum spectrum ($1/(2\pi p_T)\  dN/(d\eta
dp_T)|_{\eta=0}$ versus $p_T$) of charged particles at
$\eta=0$ in AuAu collisions at 130 GeV per nucleon in the center of
mass, for central collisions (5 \%, solid and dashed lines and filled
circles) and for minimum bias collisions (92 \%, multiplied by 0.01,
dotted and dashed-dotted
lines and open circles). Data are from PHENIX
\cite{phenix}; solid and dotted lines are results of the model
with string fusion, dashed and dashed-dotted lines with string fusion
and rescattering.

\section*{List of tables}

\noi {\bf 1.} Different particle ratios in central (10 \%)
AuAu collisions at 130 GeV
per nucleon in the center of mass in the model without string fusion or
rescattering (NF),
with string fusion (F)
and with string fusion and rescattering (FR) for particles with $p_T>
0.2$ GeV/c, compared with experimental
data \cite{brahmsratios,nuxu,brahms,phenix,star,phobosratios,starratios}.
For the centrality criteria
and kinematical cuts in the different experiments and ratios, see the
experimental references and comments in the text.

\newpage
\centerline{\bf \Large Figures:}

\newpage

\newpage

\newpage

\newpage
\centerline{\bf \Large Tables:}

\vskip 1cm

\begin{table}[htb]
\begin{center}
\begin{tabular}{cccccccc}
\hline\hline
Ratio & NF&F&FR & BRAHMS & PHENIX & PHOBOS & STAR  \\
\hline \hline
$\bar {\rm p}$/p& 0.81&0.85&0.80 & 0.64$\pm$0.04 &
0.64$\pm$0.01
& 0.60$\pm$0.04 & 0.65$\pm$0.01 \\
($\eta\sim 0$) &&& & $\pm$0.06 ($y\sim 0$)
&$\pm$0.07 & $\pm$0.06& $\pm$0.07\\ \hline
$\bar {\rm p}$/p &0.38&0.50&0.38  &0.41$\pm$0.04& &
& \\
($y\sim 2$) &&& & $\pm$0.06& & & \\ \hline
$K^-/K^+$ &0.92&0.97&0.96        & & &
0.91$\pm$0.07&
\\
($\eta\sim 0$) &&& & & & $\pm$0.06& \\ \hline
$\pi^-/\pi^+$  &1.02& 1.02&1.01       & & &
1.00$\pm$0.01& \\
($\eta\sim 0$) &&& & & & $\pm$0.02& \\ \hline
\hline
\end{tabular}
\end{center}
\caption{}
\label{tab1}
\end{table}

\end{document}